\documentstyle[11pt,twoside,epsfig,rotate,feynman,hhline]{article}

\newlength{\dinwidth}
\newlength{\dinmargin}
\newcommand{\degree}{\rm \circ}

\def\gsim{\,\lower.25ex\hbox{$\scriptstyle\sim$}\kern-1.30ex%
\raise 0.55ex\hbox{$\scriptstyle >$}\,}
\def\lsim{\,\lower.25ex\hbox{$\scriptstyle\sim$}\kern-1.30ex%
\raise 0.55ex\hbox{$\scriptstyle <$}\,}

\newcommand{\GeV}{\,\mbox{GeV}}

\newcommand{\xgrec}{$x_\gamma^{\rm rec}$}
\newcommand{\xgrecw}{x_\gamma^{\rm rec}}
\newcommand{\xprec}{$x_{\rm p}^{\rm rec}$}
\newcommand{\xprecw}{x_{\rm p}^{\rm rec}}
\newcommand{\yrec}{$y^{\rm rec}$}
\newcommand{\yrecw}{y^{\rm rec}}
\newcommand{\thetarecw}{\theta^{*,{\rm rec}}}

\begin{document}
%
%
\newcommand{\wgp}{$W_{\gamma p}$ }
\newcommand{\mx}{$M_X$} 
\newcommand{\epcs}{$ep$ cross section}
\newcommand{\pgf}{photon gluon fusion}
\newcommand{\jpsi}{$J/\psi$}
\newcommand{\jpsiw}{J/\psi} 
\newcommand{\wgpw}{W_{\gamma p}} 
\newcommand{\ep}{positron proton}
\newcommand{\ee}{$e^+e^-$} 
\newcommand{\mm}{$\mu^+\mu^-$}

\begin{titlepage}
\begin{flushleft}
{\tt DESY 97-164}\hfill {\tt ISSN 0418-9833} \\
\end{flushleft}
\vspace*{3.0cm}
\begin{center}\begin{LARGE}
{\bf Measurement of the Inclusive Di-Jet Cross Section in Photoproduction
and Determination of an Effective Parton Distribution in the Photon}\\
\vspace*{2.5cm}
H1 Collaboration \\
\vspace*{4.5cm}
\end{LARGE}
{\bf Abstract}
\begin{quotation}
\noindent
The double-differential inclusive di-jet cross section in photoproduction 
processes is measured with the H1 detector at HERA.
The cross section is determined as a function of
the average transverse jet energy $E_T^{\rm jets}$
for ranges of the fractional energy $x_\gamma^{\rm jets}$
of the parton from the photon side.
An effective leading order parton distribution in the photon is 
determined at large parton fractional energies for scales between 
\mbox{$80<p_T^2<1250$\,GeV$^2$}.
The measurement is compatible with the logarithmic scale
dependence that is predicted by perturbative QCD.
\end{quotation}
\vspace*{2.0cm}
\vfill
\cleardoublepage
\end{center}
\end{titlepage}

\begin{sloppy}
\noindent
 C.~Adloff$^{35}$,                
 S.~Aid$^{13}$,                   
 M.~Anderson$^{23}$,              
 V.~Andreev$^{26}$,               
 B.~Andrieu$^{29}$,               
 V.~Arkadov$^{36}$,               
 C.~Arndt$^{11}$,                 
 I.~Ayyaz$^{30}$,                 
 A.~Babaev$^{25}$,                
 J.~B\"ahr$^{36}$
% ...

\noindent
% ...
$ ^2$ III. Physikalisches Institut der RWTH, Aachen, Germany$^ a$ \\
% ...
$ ^{11}$ DESY, Hamburg, Germany$^a$ \\
% ...
$ ^{13}$ II. Institut f\"ur Experimentalphysik, Universit\"at Hamburg,
                                     Hamburg, Germany$^ a$  \\
% ...
$ ^{23}$ Physics Department, University of Manchester,
                                          Manchester, UK$^ b$\\
% ...
$ ^{25}$ Institute for Theoretical and Experimental Physics,
                                                 Moscow, Russia \\
 $ ^{26}$ Lebedev Physical Institute, Moscow, Russia$^ {f,k}$ \\
% ...
$ ^{29}$ LPNHE, Ecole Polytechnique, IN2P3-CNRS,
                             Palaiseau, France \\
 $ ^{30}$ LPNHE, Universit\'{e}s Paris VI and VII, IN2P3-CNRS,
                              Paris, France \\
% ...
$ ^{35}$ Fachbereich Physik, Bergische Universit\"at Gesamthochschule
               Wuppertal, Wuppertal, Germany$^ a$ \\
 $ ^{36}$ DESY, Institut f\"ur Hochenergiephysik,
                              Zeuthen, Germany$^ a$\\
% ...
$ ^a$ Supported by the Bundesministerium f\"ur Bildung, Wissenschaft,
        Forschung und Technologie, FRG,
        under contract numbers 7AC17P, 7AC47P, 7DO55P, 7HH17I, 7HH27P,
        7HD17P, 7HD27P, 7KI17I, 6MP17I and 7WT87P \\
 $ ^b$ Supported by the UK Particle Physics and Astronomy Research
       Council, and formerly by the UK Science and Engineering Research
       Council \\
% ...
$ ^f$ Supported by the Deutsche Forschungsgemeinschaft \\
% ...
$ ^k$ Supported by Russian Foundation for Basic Researches 
       grant no. 96-02-00019 \\

\end{sloppy}
\clearpage
\section{Introduction}

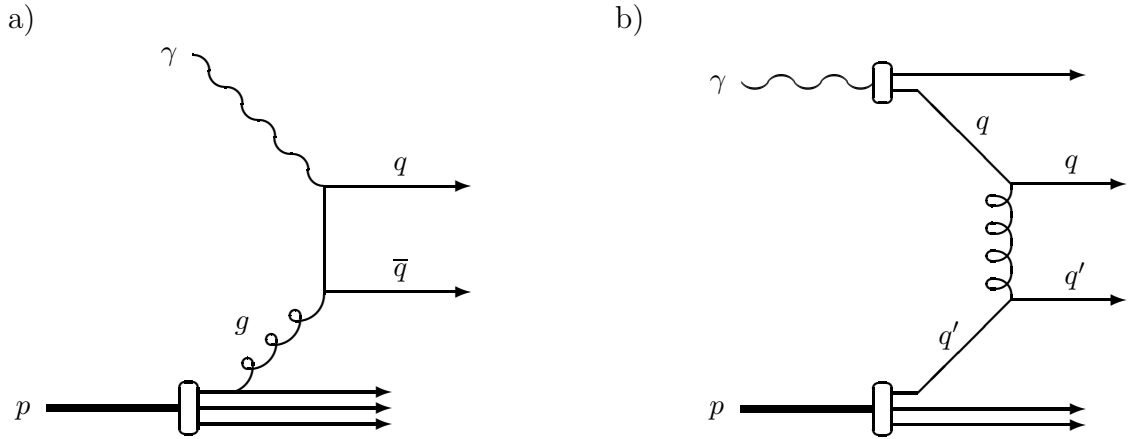
\begin{figure}[b]
\setlength{\unitlength}{0.01pt}
\begin{picture}(45000,16000)
\put(-1000,0){
\begin{picture}(20000,16000)
\THICKLINES  
\put(1000,16000){\large a)}

\drawline\fermion[\N\REG](13000,6000)[4000]
\drawline\fermion[\E\REG](\pbackx,\pbacky)[5200]
\drawarrow[\E\ATBASE](\pbackx,\pbacky)
\global\advance \pmidy by 600
\put(\pmidx,\pmidy){$q$}
\drawline\photon[\NW\REG](\pfrontx,\pfronty)[8]
\global\advance \pbackx by -1200
\global\advance \pbacky by -200
\put(\pbackx,\pbacky){$\gamma$}
\drawline\fermion[\E\REG](13000,6000)[5200]
\drawarrow[\E\ATBASE](\pbackx,\pbacky)
\global\advance \pmidy by 600
\put(\pmidx,\pmidy){$\overline{q}$}
\drawline\gluon[\SW\REG](\pfrontx,\pfronty)[3]
\global\advance \pmidx by -1500
\global\advance \pmidy by 500
\put(\pmidx,\pmidy){$g$}
\drawline\fermion[\E\REG](\pbackx,\pbacky)[6000]
\drawarrow[\E\ATBASE](\pbackx,\pbacky)
\drawline\fermion[\W\REG](\pfrontx,\pfronty)[1000]
\global\advance \pbacky by -600
\drawline\fermion[\E\REG](\pbackx,\pbacky)[7000]
\drawarrow[\E\ATBASE](\pbackx,\pbacky)
\global\advance \pfronty by -600
\drawline\fermion[\E\REG](\pfrontx,\pfronty)[7000]
\drawarrow[\E\ATBASE](\pbackx,\pbacky)
\global\advance \pfrontx by -350
\global\advance \pfronty by 600
\put(\pfrontx,\pfronty){\oval(700,2000)}
\global\advance \pfrontx by -350
\linethickness{1mm}
\drawline\fermion[\W\REG](\pfrontx,\pfronty)[5000]
\global\advance \pbackx by -1200
\global\advance \pbacky by -200
\put(\pbackx,\pbacky){$p$}
\thinlines
\THICKLINES

\end{picture}
}
\put(25000,-2300){
\begin{picture}(20000,16000)
\put(-2000,18300){\large b)}
\THICKLINES  

\drawline\gluon[\N\FLIPPED](13000,8000)[4]
\drawline\fermion[\NW\REG](\pbackx,\pbacky)[5000]
\global\advance \pmidx by 400
\global\advance \pmidy by 300
\put(\pmidx,\pmidy){$q$}
\drawline\fermion[\W\REG](\pbackx,\pbacky)[1000]
\global\advance \pbacky by 600
\drawline\fermion[\E\REG](\pbackx,\pbacky)[7000]
\drawarrow[\E\ATBASE](\pbackx,\pbacky)
\global\advance \pbackx by 1000
\global\advance \pbacky by -460
\global\advance \pfrontx by -350
\global\advance \pfronty by -300
\put(\pfrontx,\pfronty){\oval(700,1500)}
\global\advance \pfrontx by -350
\drawline\photon[\W\REG](\pfrontx,\pfronty)[5]
\global\advance \pbackx by -1200
\global\advance \pbacky by -200
\put(\pbackx,\pbacky){$\gamma$}
\drawline\fermion[\SW\REG](\gluonfrontx,\gluonfronty)[5000]
\global\advance \pmidx by -1000
\global\advance \pmidy by 100
\put(\pmidx,\pmidy){$q'$}
\drawline\fermion[\W\REG](\pbackx,\pbacky)[1000]
\global\advance \pbacky by -600
\drawline\fermion[\E\REG](\pbackx,\pbacky)[7000]
\drawarrow[\E\ATBASE](\pbackx,\pbacky)
\global\advance \pfronty by -600
\drawline\fermion[\E\REG](\pfrontx,\pfronty)[7000]
\drawarrow[\E\ATBASE](\pbackx,\pbacky)
\global\advance \pbackx by 1000
\global\advance \pbacky by 60
\global\advance \pfrontx by -350
\global\advance \pfronty by 600
\put(\pfrontx,\pfronty){\oval(700,2000)}
\global\advance \pfrontx by -350
\linethickness{1mm}
\drawline\fermion[\W\REG](\pfrontx,\pfronty)[5000]
\global\advance \pbackx by -1200
\global\advance \pbacky by -200
\put(\pbackx,\pbacky){$p$}
\thinlines
\THICKLINES

\drawline\fermion[\E\REG](\gluonbackx,\gluonbacky)[4000]
\drawarrow[\E\ATBASE](\pbackx,\pbacky)
\global\advance \pmidy by 600
\put(\pmidx,\pmidy){$q$}
\drawline\fermion[\E\REG](\gluonfrontx,\gluonfronty)[4000]
\drawarrow[\E\ATBASE](\pbackx,\pbacky)
\global\advance \pmidy by 600
\put(\pmidx,\pmidy){$q'$}
\global\advance \pbackx by 500
\global\advance \pbacky by 1850

\end{picture}
}
\end{picture}
\caption{
Diagrams for the di-jet production in $\gamma p$ scattering: 
Example for direct~(a) and resolved~(b) photon processes.
\label{feynman}}
\end{figure}
\noindent
In lepton--proton collisions at HERA the cross section is dominated
by processes in which the lepton radiates a quasi-real photon which then
interacts with the proton.
In a small fraction of such events, jets are
formed~\cite{direct, gluon, eflow, theta, erdmann}.
In the leading-order interpretation of QCD, 
the jets result either from {\em direct} 
photon interactions with the partons of the proton (Fig.\ \ref{feynman}a)
or from {\em resolved} photon--proton processes (Fig.\ \ref{feynman}b), 
where the photon interacts with the partons of the proton via its own parton
content.
For resolved photon processes, parton densities of the photon 
$f_{i/\gamma}(x_\gamma, \mu^2)$ are defined which give the 
probability of finding a parton $i$ in the photon
carrying a fraction $x_\gamma$ of the photon's energy.
The scale, $\mu$, at which the photon is probed in the
parton--parton collision is here considered to be the
transverse momentum $p_T$ of the scattered partons.

In this paper, a measurement of the 
inclusive di-jet cross section
${\rm d}^2\sigma/({\rm d}x_\gamma^{\rm jets}{\rm d}\log(E_T^{\rm jets})^2)$
is presented as a function of the observed parton fractional energy 
$0.1<x_\gamma^{\rm jets}<1$ and of the squared average
jet transverse energy $100<(E_T^{\rm jets})^2<2500$\,GeV$^2$.
Throughout the paper we denote observed hadronic variables that 
were corrected for detector effects with the suffix `jet', 
reconstructed quantities from the detector with the suffix `rec',
and leave variables referring to the partons without an additional suffix.
All transverse energies are determined with respect to the
$ep$ beam axis.

The measurement is compared with QCD calculations of the cross section
which require parton distributions of the proton and the photon as input.
In the kinematic range relevant to this analysis, the parton distributions 
of the proton are well measured.
The largest uncertainty in the results of the calculations comes 
from the quark and gluon densities of the photon.
The quark density functions used in the calculations
were determined from measurements of the photon structure function $F_2^\gamma$ 
in deep inelastic electron--photon scattering at
$e^+e^-$ colliders.
These experiments covered the kinematic range
$0.05<x_\gamma<0.9$ and $\mu^2=Q^2<400$\,GeV$^2$
where $Q^2$ is the virtuality of the photon which probes the
quasi-real photon (PETRA~\cite{pluto,jade,tasso,hermann, berger},
PEP~\cite{tpc2g},
TRISTAN~\cite{amy, topazf2} and LEP~\cite{delphi, opal}).
In contrast to these $F_2^\gamma$ measurements, the photoproduction of jets
is sensitive to both the quark and gluon content of the photon
already in leading order.
A first measurement of the gluon distribution in the photon has been
made in the range $0.04<x_\gamma<1$ 
at the scale $p_T^2=75$\,GeV$^2$ \cite{gluon}.
The comparison of the measured di-jet cross section with the QCD calculations 
provides
both a test of the universality of the parton densities in the photon and
gives new information on the parton densities.

In addition to predicting the parton--parton scattering processes
between the photon
and the proton, perturbative QCD predicts characteristic features of
the quark density $f_{\rm q/\gamma}$ in the photon~\cite{witten}.
The general characteristics are given by the DGLAP 
evolution equation for the photon
\begin{equation}
\frac{{\rm d}f_{\rm q/\gamma}(x_\gamma,\mu)}
     {{\rm d}\ln{(\mu^2/\Lambda_{\rm QCD}^2)}}
= a \,\alpha \, P_{{\rm q}\gamma}(x_\gamma)
+ b \, \alpha_s \, \int_{x_\gamma}^1 \frac{{\rm d} z}{z} 
\left[ P_{\rm qq}\left(\frac{x_\gamma}{z}\right)\, 
       f_{{\rm q}/\gamma}(z,\mu)\, 
   +\, P_{\rm qg}\left(\frac{x_\gamma}{z}\right)\, 
       f_{{\rm g}/\gamma}(z,\mu) \right]
\label{dglap}
\end{equation}
where $a$ and $b$ are constant, 
$\alpha$ and $\alpha_s$ are the fine structure constant and 
the strong coupling constant respectively,
$\Lambda_{\rm QCD}$ is the QCD scale parameter,
and $P_{ij}$ are the splitting functions.
While the form of the integral on the right hand side is the same as that
for hadrons, the inhomogeneous term 
$a \,\alpha \, P_{{\rm q}\gamma}$ is peculiar to the photon:
it reflects the contribution of quarks from the pointlike coupling of the 
photon to a quark--anti-quark pair.
Solving~(\ref{dglap}), the asymptotic behaviour of $f_{{\rm q}/\gamma}$
at large scale $\mu$ and medium $x_\gamma$ is predicted to 
be~\cite{witten}
\begin{equation}
f_{{\rm q}/\gamma} \sim \ln{\frac{\mu^2}{\Lambda_{\rm QCD}^2}}   \; .
\label{anomalous}
\end{equation}
The increase of the parton density with increasing scale is 
clearly different from the behaviour of the
parton density distributions of hadrons
and is referred to as the anomalous hadronic structure of the photon.

In this analysis, the QCD prediction
for the scale dependence of the quark densities in the photon is tested
using the di-jet measurements.
Since in the photoproduction of jets the incoming quarks and gluons
cannot be distinguished, an effective parton density
$\tilde{f}_\gamma$ of the photon is extracted from the data.
This
contains a combination of the quark and gluon densities and is valid 
in leading-order QCD \cite{combridge}.
The measured effective parton density is compared with different
predictions.

\section{Detector Description}

\noindent
A detailed description of the H1 detector can be found
elsewhere~\cite{H1}. 
Here we describe only those components which are used in the analysis.
The H1 central tracking system is mounted concentrically around the
beam-line and covers polar angles  
between $20^{\degree}<\theta<160^{\degree}$.
The angle $\theta$ is
measured with respect to the proton beam direction.
The plane perpendicular to the $z$ axis is termed $r-\phi$ plane. 
Measurements of the charge and momenta of charged particles are
provided by two coaxial cylindrical drift chambers 
(central jet chambers, CJC)~\cite{CJC}.
At two radial positions are placed a drift chamber,
which provides accurate measurement of the $z$ coordinate of charged tracks,
and a multiwire proportional chamber (MWPC), which allows triggering on
those tracks.
These are located within the inner CJC and between the two jet chambers.
The central tracking system is complemented by a forward
tracking system which covers polar angles 
$7^{\degree} \lsim\theta\lsim 25^{\degree}$.
In the present analysis the tracking detectors are used to define
the vertex position along the beam axis and to support the 
measurement of the hadronic energy flow.

The tracking system is surrounded by a highly segmented liquid argon (LAr) 
sampling 
calorimeter~\cite{calo} with an inner electromagnetic section consisting 
of lead absorber plates with a total depth of 20 to 30 radiation lengths 
and an outer hadronic section with steel absorber plates.
The LAr calorimeter covers polar angles between
$4^{\degree}$ 
and $153^{\degree}$ and
its total depth is 
4.5 to 8 interaction lengths, depending on the polar angle.
The region $151^{\degree} \lsim \theta \lsim 176^{\degree}$
is covered by a lead scintillator calorimeter (BEMC).
 
A magnetic field of 1.15~T is produced by a superconducting
solenoid surrounding the LAr calorimeter.
The iron flux return 
yoke surrounding the superconducting solenoid is instrumented
with limited streamer tubes to provide muon identification and
measurement of energy leaking
from the LAr and BEMC calorimeters.

All calorimeters provide measurement of the hadronic energy flow.
Here, jet measurements are restricted to the range which is covered
by the LAr calorimeter.

The luminosity is measured using the radiative process 
$ep\rightarrow ep\gamma$ where the photon is detected
in a luminosity monitor.

\section{Data Selection}

\noindent
The data were taken in 1994 with the H1 detector operating at the 
lepton--proton collider HERA, where positrons of 27.5\GeV\ collide 
with protons of 820\GeV.
The integrated luminosity used for this analysis is 2.9\,pb$^{-1}$. 

The photoproduction events were selected by requiring 
that the scattered electron remains in the beam pipe,
undetected in the main part of the H1 detector
(so-called untagged events).
The photon virtuality is therefore restricted to 
$Q^2 < 4$\,GeV$^2$,
where 80\% of the events have $Q^2 < 0.1$\,GeV$^2$.

The main trigger for untagged jet events is a summed transverse energy 
in the LAr calorimeter of at least
6\,GeV
with the additional requirement of a single energetic cluster of 
at least 2\,GeV
which is validated by a local coincidence of a track 
originating from the interaction vertex.
Together with triggers that are purely based on the tracking detectors
the untagged jet events are triggered with at least 
$80\%$ efficiency over the whole kinematic range.

The fraction of the lepton beam energy $E_{\rm e}$
that is carried by the photon is denoted by $y=E_{\gamma}/E_{\rm e}$.
It was approximately determined as \yrec\ from
all reconstructed objects in the central detector,
i.e.\ calorimeter clusters supported by tracking information,
with transverse energy
$E_{T,{\rm h}}^{\rm rec}$ and pseudo-rapidity
$\eta_{\rm h}^{\rm rec}=-\ln{(\tan{\theta/2})}$ using
\begin{equation}
\yrecw = \frac{1}{2E_{\rm e}}\sum_{\rm h} E_{T,{\rm h}}^{\rm rec}
e^{-\eta_{\rm h}^{\rm rec}} \; .
\label{yjb}
\end{equation}
The reconstructed scaled energy was restricted to the range $0.2<\yrecw<0.8$.
The lower cut suppresses the contribution from
background events that do not result from $ep$ interactions.
The remaining background is at the level of $1\%$.
The upper cut reduces events from 
deep inelastic scattering at $Q^2>4\,{\rm GeV}^2$,
where the scattered electron remained unidentified,
to less than $1\%$ of the total sample.

The jets were found using a cone algorithm~\cite{snow} 
with cone size $R = 0.7$ 
and jet pseudo-rapidities in the range $-{0.5} < \eta^{\rm rec} < 2.5$
in the HERA laboratory system.
In this implementation of the cone algorithm, the jet cone position
is chosen such that the transverse energy of the jet is maximal.
The jet with the highest transverse energy is constructed first, the
jet finding is then continued using only the remaining energy flow
outside of the first jet cone.
In case of overlapping jet cones, all energy flow in the overlap
region is assigned to the jet with the higher transverse energy.

After these selection cuts, the total number of events with at least
two jets reconstructed with a transverse energy above 7\,GeV was 6499.

\section{QCD Calculation of the Di-Jet Cross Section
\label{mc}}
\noindent
In leading-order QCD the differential $ep$ cross section for
di-parton production can be written as follows:
\begin{equation}
\frac{{\rm d}^4\sigma}{{\rm d}y \; {\rm d}x_{\gamma} \; 
                       {\rm d}x_{\rm p} \; {\rm d}\!\cos\theta^*}  
= 
\frac{1}{32 \pi s_{\rm ep}} 
\,\, \frac{f_{\gamma/{\rm e}}(y)}{y} 
\,\, \sum_{ij} 
\,\, \frac{f_{i/\gamma}(x_\gamma,p_T^2)}{x_\gamma} 
\,\, \frac{f_{j/{\rm p}}(x_{\rm p},p_T^2)}{x_{\rm p}}  
\,\, |M_{ij}(\cos\theta^*)|^2  \; .
\label{jet}
\end{equation}
In this expression, $s_{\rm ep}$ denotes the squared 
lepton--proton center-of-mass energy available 
at HERA, $\sqrt{s_{\rm ep}} = 300$\,GeV.
The flux of photons
radiated off the electron with 
fractional energy $y$ is predicted by QED and is denoted by 
$f_{\gamma/{\rm e}}$\@.
The fractional energy
of the parton in the photon
is given by
 $x_\gamma$
and the parton density function of parton~$i$ in the photon by $f_{i/\gamma}$.
The corresponding variables for the proton are $x_{\rm p}$ and $f_{j/{\rm p}}$.
Note that, in contrast to the quark distribution in the photon
mentioned in (\ref{dglap}), 
$f_{i/\gamma}$ here represents all components in the
photon including the direct photon processes and the 
quarks and gluons for resolved photon processes.
The last term in (\ref{jet}) contains the distribution of the 
parton scattering angle $\theta^*$ in the parton--parton center-of-mass 
system in the form of the QCD matrix elements $M_{ij}(\cos\theta^*)$. 

The parton transverse momentum $p_T$, which is here identified
with the scale of the process, can be expressed in terms of the 
four observables in (\ref{jet}):
\begin{equation}
p_T^2 = \frac{1}{4} s_{\rm ep} \, y \, x_{\gamma} \, 
        x_{\rm p} \, \sin^2\theta^* \; .
\end{equation}

For the correction of the data, and for comparison to the measured
jet cross sections, the two event generators 
PYTHIA~\cite{pythia} and PHOJET~\cite{Ralph,Ralph2} were used. 
In both generators jet production is based on
the leading-order di-parton cross section
as given in~(\ref{jet}).

The PYTHIA 5.7 event generator~\cite{pythia}
was used
to simulate photon-proton interactions.
As well as the leading-order cross section~(\ref{jet}),
PYTHIA includes initial- and final-state parton radiation effects
which are calculated in the leading logarithmic approximation.
The strong coupling constant 
$\alpha_s$ is calculated in first order QCD using
$\Lambda_{\rm QCD}=200$\,MeV for 4 quark flavours.
Multiple parton interactions were generated in
addition to the primary parton--parton scattering.
They are calculated as leading-order QCD processes between partons from the
photon and proton remnants.
The
transverse momentum
of all parton--parton interactions was required to be above
$p_T=1.2$\,GeV\@.
This cut-off value has been found to give an
optimal description of the transverse energy
flow outside the jets~\cite{eflow}.
A good description of this energy flow is important since the
contributions of multiple parton interactions inside the jet
cones alter the jet rates considerably.
For tests of the uncertainties in the determination of the effective
parton distribution of the photon arising from the contribution of
multiple parton interactions, an event sample was generated 
with $p_T>1.4$\,GeV, which gives an energy flow that is significantly smaller
than that observed in the data.
To assess the significance of higher-order effects, another sample was
generated without initial-state parton showers. GRV-LO parameterizations 
were used throughout
for the parton distributions of the photon and the
proton~\cite{ggrv,pgrv}
and  hadronization was modelled with the LUND string
fragmentation scheme (JETSET 7.4~\cite{jetset}).

The PHOJET 1.06 event generator~\cite{Ralph,Ralph2}
is based on the two-component Dual Parton Model~\cite{DPM}.
It calculates parton--parton scattering using~(\ref{jet}).
Again, $\Lambda_{\rm QCD}$ is 200\,MeV for 4 quark flavours.
PHOJET incorporates very detailed simulations of both
multiple soft and hard parton interactions
on the basis of a unitarization scheme~\cite{unitar}.
It includes initial- and final-state hard parton radiation effects.
The lower momentum cut-off for hard parton interactions was set to
$p_T=2.5$\,GeV.
In contrast to the PYTHIA generator, small variations of this cut-off
value do not have a large influence on the results obtained
by PHOJET, due to the unitarization scheme.
For the fragmentation the LUND string concept is applied
(JETSET 7.4~\cite{jetset}).
The GRV-LO parton distribution functions for the photon and the proton
were also used here.
The effects of higher-order corrections were
checked by
generating a sample of events without hard initial-state parton showers.

For comparisons with the measured jet cross sections, 
analytic QCD calculations were provided by ref.~\cite{klasen}.
These calculations include the next-to-leading order (NLO) 
QCD matrix elements.
They use the NLO parton distribution functions GRV-HO~\cite{ggrv}
and GS96 \cite{gs} for the photon and the
CTEQ-4M distributions~\cite{cteq} for the partons of the proton.

\section{Measurement of the Di-Jet Cross Section%
\label{dijet}}
\noindent
In order to study observables closely related to the parton 
distributions in the photon, the differential di-jet cross sections
were analysed in terms of the variables
$E_T^{\rm jets}$ and $x_\gamma^{\rm jets}$.
The observable $E_T^{\rm jets}$ was calculated using the 
two jets with the highest $E_T^{\rm jet}$ in the event
\begin{equation}
E_T^{\rm jets} = \frac{E_{T}^{\rm jet1}+E_{T}^{\rm jet2}}{2}
\label{sumetcut}
\end{equation}
and was required to be above $E_T^{\rm jets} = 10$\,GeV\@.

The ratio of the difference and the sum of
the transverse energies of the jets was required to be in the range
\begin{equation}
\frac{|E_{T}^{\rm jet1}-E_{T}^{\rm jet2}|}
     {E_{T}^{\rm jet1}+E_{T}^{\rm jet2}} < 0.25 \; .
\label{deltaetcut}
\end{equation}
The cuts~(\ref{sumetcut}) and~(\ref{deltaetcut}) ensure that the transverse
energy of all jets is above $E_T^{\rm jet}=7.5$\,GeV,
without using the same $E_T^{\rm jet}$ cut-off for both jets.
In next-to-leading order analytical calculations of the di-jet cross section
this avoids introducing a dependence of the result on unphysical parameters.
%
%
%
%

The observable $x_\gamma^{\rm jets}$ was calculated
using the transverse energies and pseudo-rapidities of the two
jets and the photon energy:
\begin{equation}
x_{\gamma}^{\rm jets} 
= \frac{E_{T}^{\rm jet1} e^{-\eta^{\rm jet1}} 
      + E_{T}^{\rm jet2} e^{-\eta^{\rm jet2}}}
       {2 y E_{\rm e}}  \; .
\label{xg}
\end{equation}
Here, the interval $0.1<x_{\gamma}^{\rm jets}<1$ is considered.

The average rapidity of the two jets was required to be in the range
\begin{equation}
0 < \frac{\eta^{\rm jet1}+\eta^{\rm jet2}}{2} < 2 \; .
\label{boost}
\end{equation}
This variable is approximately the rapidity of the parton--parton center-of-mass
system in the laboratory frame of reference.
The difference in the jet pseudo-rapidities was required to be within
\begin{equation}
\vert\Delta\eta^{\rm jets}\vert < 1
\end{equation}
which corresponds to $\vert\cos{\theta^{*,{\rm jets}}}\vert<0.46$.
Together with~(\ref{boost}) this cut ensures that all jets are in a region
in which the hadronic energy is well measured in the detector.

\begin{figure}
\setlength{\unitlength}{1cm}
\begin{picture}(10.0,15.0)
\put(2.0,-0.2)
{\epsfig{file=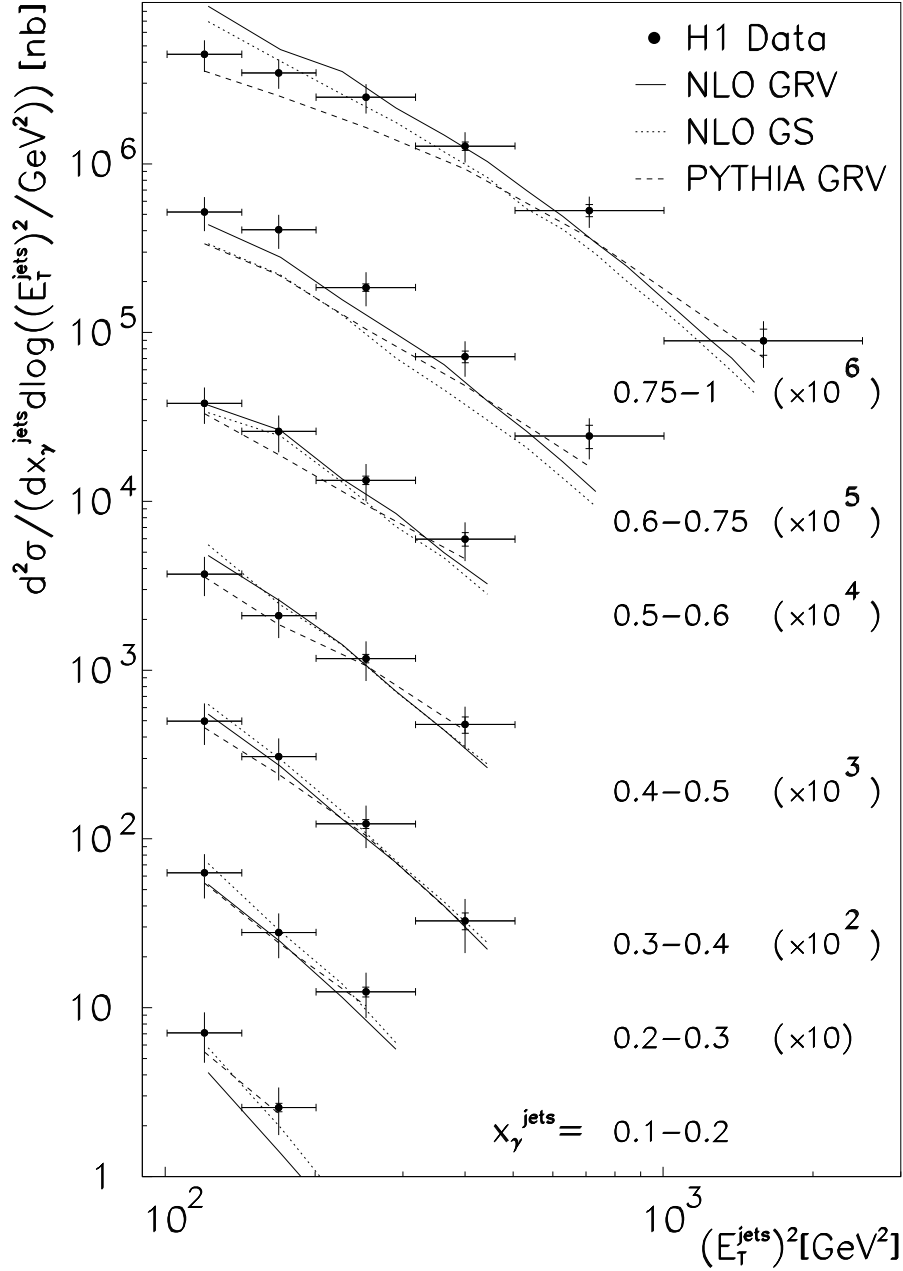, bb=45 48 520 736, width=11.8cm}}
\end{picture}
\caption{ 
The measured inclusive di-jet $ep$ cross section is shown as a function
of the squared jet transverse energy $E_T^{\rm jets}$
for ranges of the reconstructed parton fractional energy $x_\gamma^{\rm jets}$ 
(full circles).
$E_T^{\rm jets}$ is the average transverse energy of the two jets
with the highest $E_T^{\rm jet}$ in the event.
The average pseudo-rapidities of the jets were between 
$0<(\eta^{\rm jet1}+\eta^{\rm jet2})/2<2$ and their difference was
below $\vert\Delta\eta^{\rm jets}\vert=1$.
The transverse energies of the jets were restricted to the range
$|E_{T}^{\rm jet1}-E_{T}^{\rm jet2}|/
     (E_{T}^{\rm jet1}+E_{T}^{\rm jet2}) < 0.25$.
The cross section is integrated over the photon virtuality $Q^2<4$\,GeV$^2$
and the relative photon energy $0.2<y<0.83$.
The inner error bars represent the statistical errors, the outer error
bars give the statistical and systematic errors, added in quadrature.
The data are compared to the QCD simulation of the PYTHIA generator
using the GRV-LO parton distribution functions
(dashed curve)
and to analytical next-to-leading (NLO) QCD calculations
of the di-jet cross section~\cite{klasen} 
using the GRV-HO (full line) and GS96 (dotted
line) photon parton distributions.
\label{xsecpt}}
\end{figure}
\begin{table}
\begin{center}
\begin{tabular}{||lr|rc||c|c|cc|c||}
\hhline{|t:=========:t|}
 & & & & cross section & \multicolumn{4}{c||}{error} \\
 \multicolumn{2}{||c|}{$x_\gamma^{\rm jets}$} & \multicolumn{2}{c||}{$\log(E_T^{\rm jets}/{\rm GeV})^2$}
& $\frac{{\rm d}^2\sigma^{ep\rightarrow {\rm 2 jets}}}
        {{\rm d}x_\gamma^{\rm jets}
         {\rm d}\log{((E_T^{\rm jets})^2/{\rm GeV}^2)}}$ &
stat. & \multicolumn{2}{c|}{syst.} & total\\
 min  & max & $\quad$min & max & $[$nb$]$ & & E-scale & other & \\
\hline\hline
 0.10 & 0.20 & 2.00 & 2.15 & 7.42 & $\pm$0.27 & $\pm$2.00 & $\pm$0.83 & $\pm$2.19 \\
      &      & 2.15 & 2.30 & 2.62 & $\pm$0.14 & $\pm$0.71 & $\pm$0.29 & $\pm$0.78 \\
\hline
 0.20 & 0.30 & 2.00 & 2.15 & 6.31 & $\pm$0.20 & $\pm$1.58 & $\pm$0.71 & $\pm$1.74 \\
      &      & 2.15 & 2.30 & 2.83 & $\pm$0.13 & $\pm$0.71 & $\pm$0.32 & $\pm$0.79 \\
      &      & 2.30 & 2.50 & 1.23 & $\pm$0.08 & $\pm$0.31 & $\pm$0.14 & $\pm$0.35 \\
\hline
 0.30 & 0.40 & 2.00 & 2.15 & 5.06 & $\pm$0.19 & $\pm$1.16 & $\pm$0.57 & $\pm$1.31 \\
      &      & 2.15 & 2.30 & 3.00 & $\pm$0.14 & $\pm$0.69 & $\pm$0.34 & $\pm$0.78 \\
      &      & 2.30 & 2.50 & 1.27 & $\pm$0.08 & $\pm$0.29 & $\pm$0.14 & $\pm$0.33 \\
      &      & 2.50 & 2.70 & 0.30 & $\pm$0.03 & $\pm$0.09 & $\pm$0.03 & $\pm$0.10 \\
\hline
 0.40 & 0.50 & 2.00 & 2.15 & 3.81 & $\pm$0.15 & $\pm$0.80 & $\pm$0.43 & $\pm$0.92 \\
      &      & 2.15 & 2.30 & 2.19 & $\pm$0.10 & $\pm$0.46 & $\pm$0.24 & $\pm$0.53 \\
      &      & 2.30 & 2.50 & 1.16 & $\pm$0.07 & $\pm$0.24 & $\pm$0.13 & $\pm$0.28 \\
      &      & 2.50 & 2.70 & 0.46 & $\pm$0.05 & $\pm$0.10 & $\pm$0.05 & $\pm$0.12 \\
\hline
 0.50 & 0.60 & 2.00 & 2.15 & 4.05 & $\pm$0.16 & $\pm$0.77 & $\pm$0.45 & $\pm$0.91 \\
      &      & 2.15 & 2.30 & 2.71 & $\pm$0.13 & $\pm$0.51 & $\pm$0.30 & $\pm$0.61 \\
      &      & 2.30 & 2.50 & 1.27 & $\pm$0.08 & $\pm$0.24 & $\pm$0.14 & $\pm$0.29 \\
      &      & 2.50 & 2.70 & 0.60 & $\pm$0.05 & $\pm$0.11 & $\pm$0.07 & $\pm$0.14 \\
\hline
 0.60 & 0.75 & 2.00 & 2.15 & 5.11 & $\pm$0.18 & $\pm$0.84 & $\pm$0.61 & $\pm$1.06 \\
      &      & 2.15 & 2.30 & 4.15 & $\pm$0.16 & $\pm$0.69 & $\pm$0.46 & $\pm$0.84 \\
      &      & 2.30 & 2.50 & 2.00 & $\pm$0.10 & $\pm$0.33 & $\pm$0.22 & $\pm$0.41 \\
      &      & 2.50 & 2.70 & 0.75 & $\pm$0.06 & $\pm$0.12 & $\pm$0.08 & $\pm$0.16 \\
      &      & 2.70 & 3.00 & 0.27 & $\pm$0.04 & $\pm$0.04 & $\pm$0.03 & $\pm$0.07 \\
\hline
 0.75 & 1.00 & 2.00 & 2.15 & 4.32 & $\pm$0.18 & $\pm$0.54 & $\pm$0.59 & $\pm$0.82 \\
      &      & 2.15 & 2.30 & 3.43 & $\pm$0.14 & $\pm$0.42 & $\pm$0.38 & $\pm$0.59 \\
      &      & 2.30 & 2.50 & 2.50 & $\pm$0.11 & $\pm$0.31 & $\pm$0.28 & $\pm$0.43 \\
      &      & 2.50 & 2.70 & 1.25 & $\pm$0.07 & $\pm$0.16 & $\pm$0.14 & $\pm$0.22 \\
      &      & 2.70 & 3.00 & 0.54 & $\pm$0.04 & $\pm$0.07 & $\pm$0.06 & $\pm$0.10 \\
      &      & 3.00 & 3.40 & 0.09 & $\pm$0.02 & $\pm$0.02 & $\pm$0.01 & $\pm$0.03 \\
\hhline{|b:=========:b|}
\end{tabular} \\
\end{center}
\caption{
Inclusive double-differential di-jet cross section in bins of 
$x_\gamma^{\rm jets}$
and $\log(E_T^{\rm jets}/{\rm GeV})^2$.
The limits of the bins are listed
in the first four columns.
For each bin a cross section number is given together with
the statistical and systematic uncertainties.
The error induced by the 4\% uncertainty
of the hadronic calibration of the calorimeter (E-scale) is listed separately.
\label{dijettable}}
\end{table}

In Fig.~\ref{xsecpt} the $ep$ double-differential di-jet cross section is 
shown as a function of the transverse energy $E_T^{\rm jets}$
in bins of the fractional parton energy $x_\gamma^{\rm jets}$. 
The event kinematics corresponds to corrected photon fractional energies 
in the range
\begin{equation}
0.2 < y < 0.83
\end{equation}
and photon virtualities of
\begin{equation}
Q^2 < 4 \, {\rm GeV}^2 \; .
\end{equation}
The di-jet cross section was corrected for detector effects
using an unfolding procedure~\cite{dagostini}.
The procedure requires a 4-dimensional correlation function 
between the variables reconstructed in the detector
$(x_{\gamma}^{\rm rec}, E_T^{\rm rec})$ and
the corrected observables referring to the hadronic final state
$(x_{\gamma}^{\rm jets}, E_T^{\rm jets})$.
This correlation was calculated using the PYTHIA generator together
with a detailed simulation of the detector.
The procedure modifies the di-jet cross section of the PYTHIA
calculation without detector effects so as to ensure that the 
predicted di-jet distributions as a function of
$(x_{\gamma}^{\rm rec}, E_T^{\rm rec})$
agree with the data.
Further details of the correction procedure can be found in~\cite{rick}.
The inner error bars reflect the statistical errors. 
The outer error bars shown are the statistical and systematic errors added in 
quadrature. 
The main contribution to the systematic error results from a 4\%~uncertainty 
in the knowledge of the calorimeter hadronic energy scale 
(Table~\ref{dijettable}).
Variation of the unfolding parameters
affects the measured cross section by less than $10\%$.
The uncertainty in the determination of the trigger efficiency 
using redundant triggers
is below $10\%$.
The luminosity measurement is accurate to $1.5\%$.

The measured cross section is compared to the simulation
of the PYTHIA generator (dashed curve)
using the GRV-LO parton distribution functions for the 
photon and the proton~\cite{ggrv,pgrv}. 
This calculation includes the leading-order QCD matrix elements
together with higher-order QCD effects which are simulated by a
parton shower mechanism, multiple parton interactions
and hadronization.
All $x_\gamma^{\rm jets}$ ranges are dominated
by resolved photon processes, except for the highest region,
$x_\gamma^{\rm jets}>0.75$, where the direct photon processes
give the dominant contribution.
The PHOJET generator, using the same parton distribution functions
and the same choice of the scale, gives results that are compatible
with the PYTHIA calculation to within $20\%$ (not shown).
Also shown are two analytical calculations of the di-jet cross
section in next-to-leading order QCD~\cite{klasen} which use
the GRV-HO~\cite{ggrv} (full curve) and GS96~\cite{gs} (dotted curve)
photon parton distribution functions.
In both calculations the CTEQ-4M~\cite{cteq} parameterization of
the proton parton distribution is used.
In these calculations the jet finding is based on the three final
state partons.
Since fragmentation effects and
underlying event energy from multiple parton interactions 
are not included,
deviations from the measured cross sections are expected.
Multiple parton interactions
increase the jet energy and therefore the jet cross
section.
The influence on the cross section has been estimated to be below 40\% at small
$x_\gamma^{\rm jets}\approx 0.1$ and simultaneously small $E_T^{\rm jets}$,
and is negligible at
$x_\gamma^{\rm jets}\approx 1$ or $E_T^{\rm jets}>15\,{\rm GeV}$.
The influence of the fragmentation process is maximal in the region of
low $E_T^{\rm jets}$ and
raises the jet cross section
by at most 20\%~\cite{heraws}.

All calculations give a satisfactory description of the data overall,
except for some deficiencies in the lowest $x_\gamma^{\rm jets}$ range
and the two highest $x_\gamma^{\rm jets}$ ranges.
In these regions, the resolved photon contribution is not well constrained
from measurements of the photon structure function $F_2^\gamma$ at
$e^+e^-$~colliders.
The two GRV-LO and the GS96 parton distribution functions each give a good
description of the $F_2^\gamma$ measurements.
The successful application of these parton distributions
in the calculation of the di-jet cross section in $ep$ scattering
supports the universality of the photon parton distributions.
However, in detail, the differences between the measurement and these
calculations show that the jet data give new constraints on the parton 
distribution functions,
especially in the region at large $x_\gamma^{\rm jets}$
and high $E_T^{\rm jets}$\@.

\section{Effective Parton Distribution of the Photon}

\noindent
In order to extract the parton distributions of the photon from
the data, $f_{i/\gamma}$ in (\ref{jet})
needs to be 
disentangled from the sum over the different parton processes.
This is not possible without further assumptions.

Here we follow the approximation procedure developed
in~\cite{combridge}.
The method is based on the leading-order matrix elements for the
partonic scattering processes
$qq'\rightarrow qq'$, $qg \rightarrow qg$,
and $gg \rightarrow gg$ which give the dominant
contribution to the total resolved photon di-jet cross section.
The shapes of the angular distributions of the scattered partons
are similar for all of these processes.
The squared matrix elements differ in rate by the ratio of the 
colour factors $C_A/C_F=9/4$.
The approximation uses only one Single Effective 
Subprocess matrix element $M_{\rm SES}$
and combines the quark and gluon densities into new effective parton
density functions
\begin{eqnarray}
\tilde{f}_{\gamma}(x_\gamma, p_T^2) & \equiv &
\sum_{\rm n_f} \left(f_{{\rm q}/\gamma}(x_\gamma, p_T^2)
               +f_{\overline{\rm q}/\gamma}(x_\gamma, p_T^2)\right) +
\frac{9}{4} \,  f_{{\rm g}/\gamma}(x_\gamma, p_T^2)
\label{fgamma}
\;\;\;\;{\rm and} \\
\tilde{f}_{\rm p}(x_{\rm p}, p_T^2) & \equiv &
\sum_{\rm n_f} \left(f_{{\rm q}/{\rm p}}(x_{\rm p}, p_T^2)
               +f_{\overline{\rm q}/{\rm p}}(x_{\rm p}, p_T^2)\right) +
\frac{9}{4} \,  f_{{\rm g}/{\rm p}}(x_{\rm p}, p_T^2)
\; .
\end{eqnarray}
The sums run over the quark flavours.
The product 
$\tilde{f}_{\gamma} \tilde{f}_{\rm p} \vert M_{\rm SES} \vert^2$
then replaces the
resolved photon contributions to the sum
$\sum_{ij} f_{i/\gamma} f_{j/{\rm p}} \vert M_{ij}\vert^2$
of equation (\ref{jet}):
\begin{equation}
\frac{{\rm d}^4\sigma}{{\rm d}y \; {\rm d}x_{\gamma} \; 
                       {\rm d}x_{\rm p} \; {\rm d}\!\cos\theta^*}  
\approx \frac{1}{32 \pi s_{\rm ep}} 
\,\, \frac{f_{\gamma/{\rm e}}(y)}{y} 
\,\, \frac{\tilde{f}_{\gamma}(x_\gamma,p_T^2)}{x_\gamma} 
\,\, \frac{\tilde{f}_{\rm p}(x_{\rm p},p_T^2)}{x_{\rm p}}  
\,\, \vert M_{\rm SES}(\cos\theta^*) \vert^2 \; .
\label{ses}
\end{equation}
Here $\tilde{f}_{\gamma}(x_\gamma,p_T^2)$ can be directly determined from the
measurement of the di-parton cross section.
In the kinematic range which is relevant to this analysis
the accuracy of the $9/4$ weight in (\ref{fgamma}) was checked
with a PYTHIA calculation of the di-parton cross section based 
on (\ref{jet}).
The relative contributions 
of the quark and gluon induced di-jet
events agree with this weight to better than $5\%$
in the range $0.2<x_\gamma<0.7$ considered here.

\subsection{Test of the Factorization of the Di-Jet Cross Section}
\begin{figure}[!ht]
\setlength{\unitlength}{1cm}
\begin{picture}(15.5,13.2)
\put(-0.3,-0.8)
{\epsfig{file=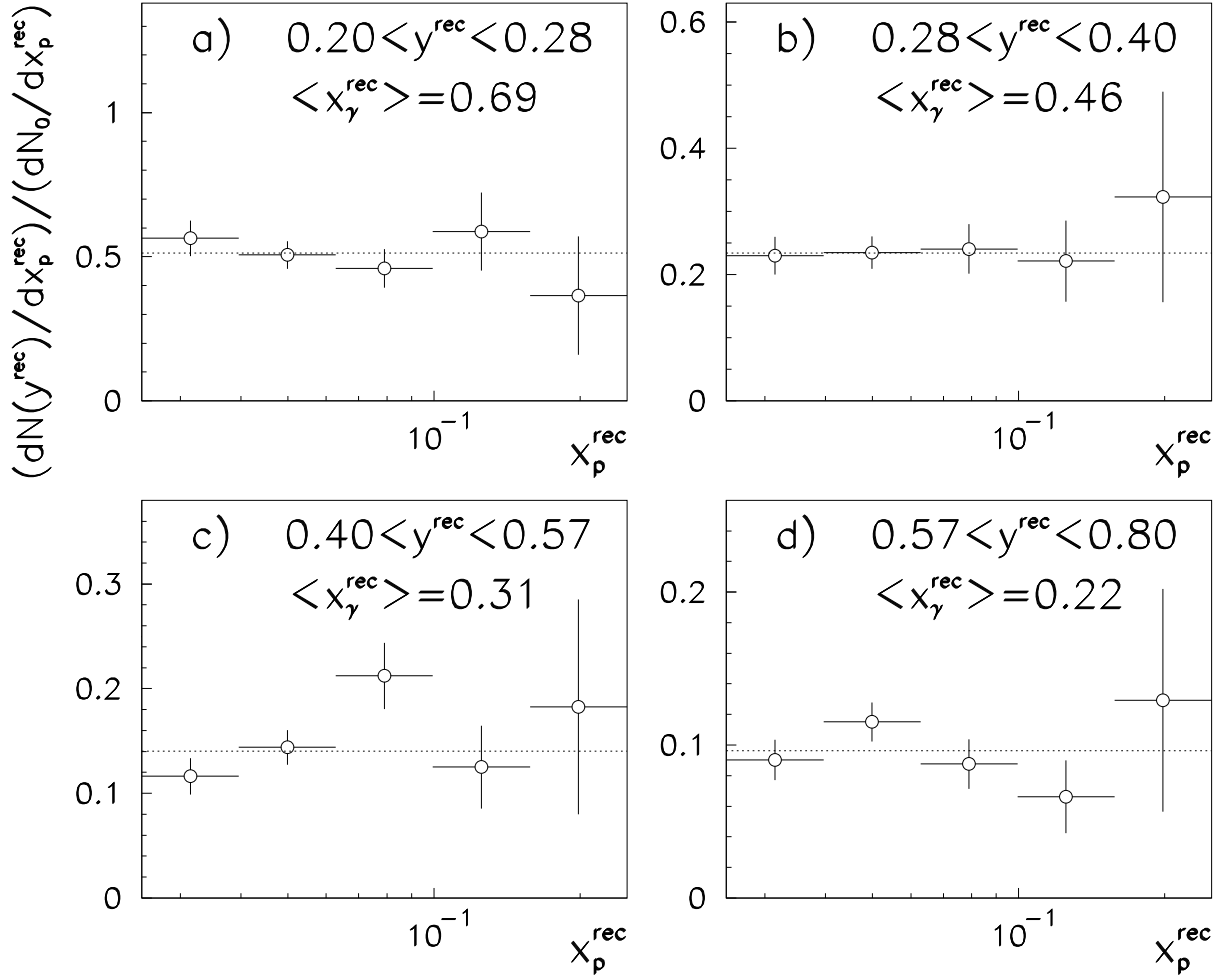, width=17.2cm}}
\end{picture}
\caption{
The distributions of the reconstructed fractional energy \xprec\ of the parton 
from the proton
side are shown in four bins of the photon fractional energy \yrec.
All distributions were normalized to the
\xprec\ distribution of all events
in the whole interval $0.2<\yrecw<0.8$, ${\rm d}N_0/{\rm d}\xprecw$\@.
Only statistical errors are shown.
The dotted line represents a fit to the ratio assuming a flat distribution.
\label{factor}}
\end{figure}

Before the effective parton density $\tilde{f}_{\gamma}$
of the photon is extracted, 
it is shown that the factorization of the di-jet cross section 
(\ref{ses}) into effective parton densities is compatible with the
measured data.
If the factorization approximation is good,
the shape of the parton fractional energy
distribution \xprec\ from
the proton side should be independent of the parton fractional energy
from the photon side \xgrec.
In order to test this hypothesis, the four dimensions of the
reconstructed di-jet distribution 
$N( \yrecw, \xgrecw, \xprecw, \thetarecw)$ 
are reduced to two dimensions, using the two independent
observables \xprec\ and \yrec\ by applying the requirements
$\thetarecw \approx 90^\circ$ ($\vert\Delta\eta^{\rm rec}\vert<1$)
and $\yrecw \, \xgrecw \approx$ constant 
($0.1 < \yrecw \, \xgrecw < 0.2$).
The latter cut implies that the energy which enters the
parton--parton scattering process from the electron side is constant.
\xprec\ is calculated from the reconstructed jet energies and
pseudo-rapidities using
$\xprecw= \left(E_{T,1}^{\rm rec} \, e^{\eta_1^{\rm rec}}
              + E_{T,2}^{\rm rec} \, e^{\eta_2^{\rm rec}}\right)/
          \left(2 E_{\rm p}\right)$.
In Fig.~\ref{factor}, the \xprec\ distribution is 
shown in 4 different \yrec\ intervals.
All distributions were normalized to the data sample
covering the full \yrec\ interval, $0.2<\yrecw<0.8$.
The measured ratios are compatible with having the same shape in all four 
\yrec\ bins.

Since the average parton fractional energy 
$\langle \xgrecw \rangle$
of the four \xprec\ distributions varies between 
$\langle \xgrecw \rangle=0.22$ and
$\langle \xgrecw \rangle=0.69$,
the observed \xprec\ distributions are also independent of \xgrec.
Therefore, within the precision of the data, factorization holds in 
(\ref{ses}) and so the ansatz used to extract an
effective
parton distribution for the photon is meaningful.
\subsection{Effects Influencing the Correction
to Leading Order Parton Variables}
In order to correct from the jet observables to the parton variables
three effects have to be taken into account.
These are multiple parton interactions, higher-order QCD effects
and fragmentation effects.

In the generators used here, the fragmentation is simulated
using the
well tuned
Lund string fragmentation model (JETSET 7.4).

Multiple parton interactions lead to an additional energy flow
in the event which affects the jet rates. These effects were studied
in a previous analysis using the transverse energy flow outside the
jets and energy-energy correlations. These are well simulated, for example,
by the PYTHIA generator~\cite{eflow}.

Higher-order QCD effects
can be studied by looking at multi-jet production.
The relative contribution of events with three or more reconstructed jets above
a transverse jet energy of 7\,GeV amounts to 8\% of the total event sample.
This contribution is well described by the PYTHIA simulation.

Further results of higher-order QCD effects are
 an imbalance between the 
transverse energies of the two highest transverse energy jets and a
deviation from a back-to-back configuration in the azimuth.
For this study,
the region of reconstructed parton fractional energies 
above $\xgrecw=0.4$ is considered
in order to minimize contributions of multiple parton interaction effects
~\cite{eflow}.
The missing total transverse energy in photoproduction events is small
and was here required to be below
$E_{T,{\rm miss}}^{\rm rec}=5$\,GeV to ensure that the transverse jet energies
were well measured.

\begin{figure}[!ht]
\setlength{\unitlength}{1cm}
\begin{picture}(14.0,6.5)
\put(0,0)
{\epsfig{file=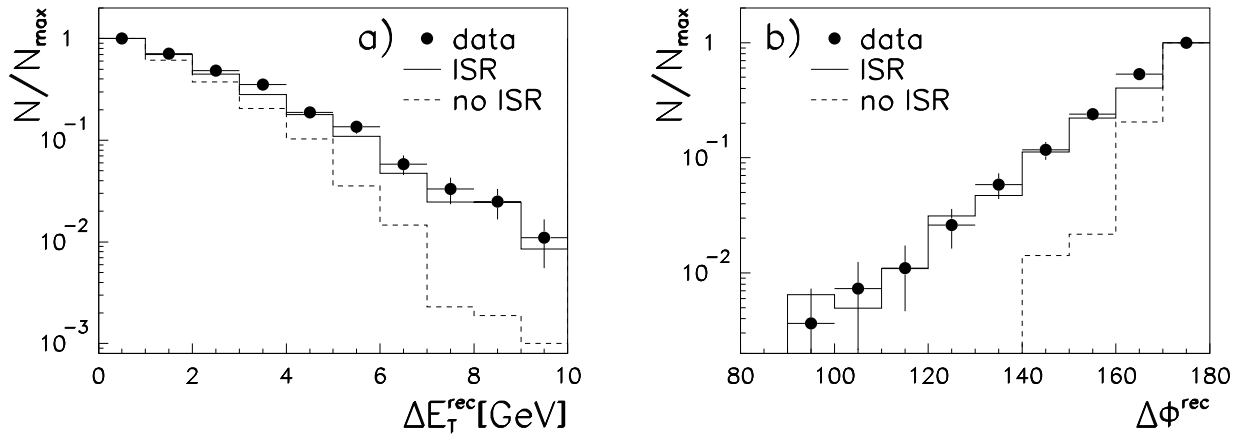, bb=19 20 710 300, width=\textwidth}}
\end{picture}
\caption{
a) The shape of the distribution of the
uncorrected jet transverse energy imbalance
$\Delta E_T^{\rm rec}=\vert E_{T,1}^{\rm rec}-E_{T,2}^{\rm rec}\vert$ is shown.
The distribution is normalized to the maximum number of entries $N_{\rm max}$
in a single bin.
The reconstructed parton fractional energy was required to be $\xgrecw>0.4$\@.
The full histogram shows the prediction of the PYTHIA generator
which includes hard initial-state parton radiation effects (ISR).
The dashed histogram shows the calculation of the PHOJET generator
in a version without hard initial-state parton radiation (no ISR).
Both calculations include a detailed simulation of the detector effects.
b) The shape of the uncorrected distribution of the
azimuthal difference between the two jets
$\Delta \phi^{\rm rec}=\vert \phi_1^{\rm rec}-\phi_2^{\rm rec}\vert$
is shown.
The histogram assignments are as in a).
\label{ips}}
\end{figure}

In Fig.~\ref{ips}a, the shape of the transverse energy imbalance
$\Delta E_T^{\rm rec}=\vert E_{T,1}^{\rm rec}-E_{T,2}^{\rm rec}\vert$
between the two jets 
with the highest $E_T^{\rm rec}$ is shown.
Fig.~\ref{ips}b shows the azimuthal difference between the two jets
$\Delta \phi^{\rm rec}=\vert \phi_1^{\rm rec}-\phi_2^{\rm rec}\vert$.
The shape of the $\Delta E_T^{\rm rec}$ and $\Delta \phi^{\rm rec}$
distributions is described by
a PYTHIA calculation which includes hard initial-state parton showers
(full histogram in Fig.~\ref{ips}).
This calculation also gives a good description of the 
$\Delta E_T^{\rm rec}$ and $\Delta \phi^{\rm rec}$ 
distributions at $\xgrecw<0.4$ (not shown).
The dashed histogram represents a calculation of the PHOJET
generator in a version that does not include hard initial-state parton
radiation effects: this calculation gives too small $\Delta E_T^{\rm rec}$
and too large $\Delta \phi^{\rm rec}$.
We conclude that
higher-order QCD effects are well modelled by
parton showers.

\subsection{Extraction of the Effective Parton Distribution Function}
For the extraction of the effective photon parton distribution function,
$\tilde{f}_{\gamma}$, the measured di-jet cross section was corrected to the
level of the leading-order di-parton cross section using the same unfolding
procedure~\cite{dagostini} 
as applied in the analysis of the jet cross section described above.
In this case the correlations between 
$(x_{\gamma}^{\rm jets}, E_T^{\rm jets})$
and the parton variables $(x_\gamma, p_T)$
are used to correct for the fragmentation, higher-order QCD effects and
underlying event energy effects.
As discussed in the previous paragraph, all these effects are well modelled
by the PYTHIA generator.
The model dependence of the corrections was checked
using the different generated event samples
described in Sect.~\ref{mc}.
Comparison of the di-parton cross section 
from data and a PYTHIA calculation using (\ref{jet}) together with 
the GRV-LO parton distributions and with $p_T$ as the scale
then gives the effective parton distribution of the photon in the data:
\begin{equation}
\tilde{f_\gamma}^{\rm DATA} = \, \tilde{f_\gamma}^{\rm GRV-LO}
\, \times \,
\frac{{\rm d}^2\sigma^{\rm DATA}/({\rm d}x_\gamma {\rm d}\log{p_T^2})}
{{\rm d}^2\sigma^{\rm PYTHIA, GRV-LO}/({\rm d}x_\gamma {\rm d}\log{p_T^2})}
\ \ .\end{equation}

\begin{figure}
\setlength{\unitlength}{1cm}
\begin{picture}(12.0,17.0)
\put(2.0,-0.3)
{\epsfig{file=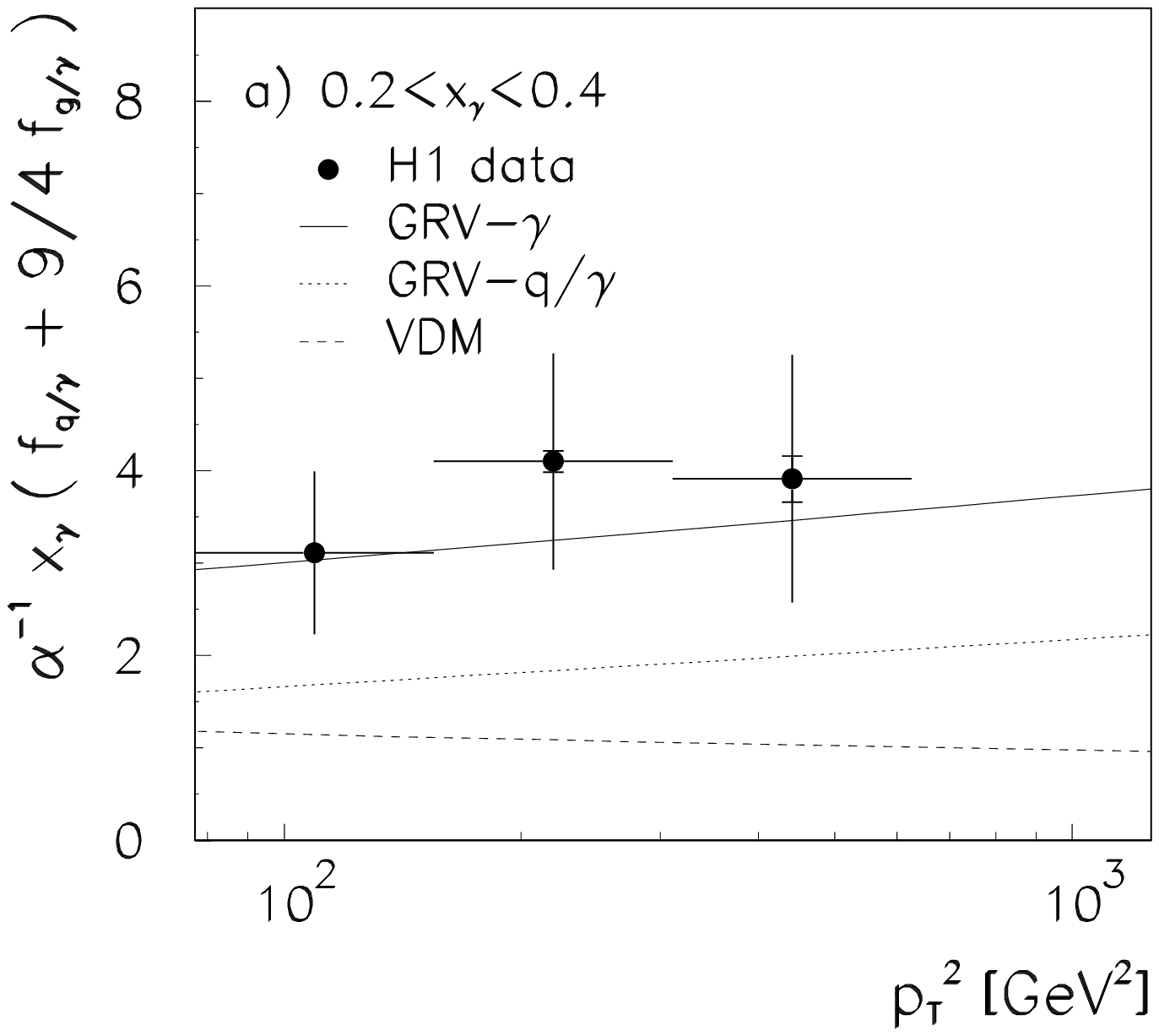, width=11cm}}
\put(2.0,9.0)
{\epsfig{file=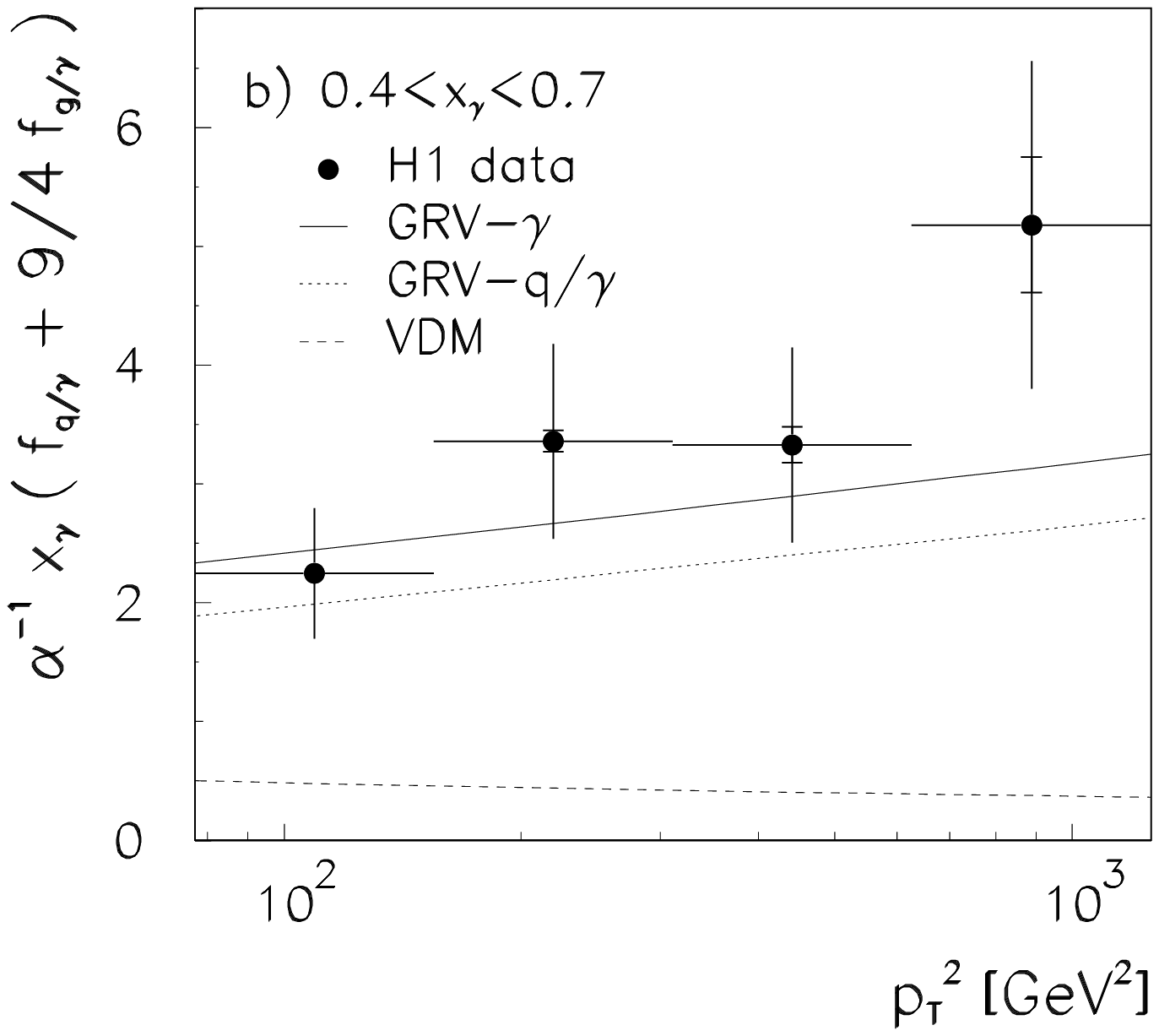, width=11cm}}
\end{picture}
\caption{
The leading order effective parton distribution of the photon 
$x_\gamma \tilde{f}_{\gamma}(x_\gamma,p_T^2)=
x_\gamma(f_{{\rm q}/\gamma}(x_\gamma,p_T^2) + 9/4\,f_{{\rm g}/\gamma}(x_\gamma,p_T^2))$
is shown as a function of the squared parton transverse momentum $p_T^2$.
The $f_{{\rm q}/\gamma}$ represents here the sum over all quarks and anti-quarks and
$f_{{\rm g}/\gamma}$ is the gluon density.
The distribution was divided by the fine structure constant~$\alpha$
and averaged over parton fractional energies in the ranges
a) $0.2<x_\gamma<0.4$ and b) $0.4<x_\gamma<0.7$.
The inner error bars represent the statistical errors, the outer error
bars give the statistical and systematic errors, added in quadrature.
The data are compared to the effective parton distribution of 
the photon which includes the pointlike coupling of the photon to quarks
(full curve) and to a vector meson dominance (VDM) ansatz for the photon
(dashed curve).
The dotted curve shows the quark part of the effective parton distribution.
All three curves were
calculated using the GRV-LO parton distribution functions.
\label{sf}}
\end{figure}

In Fig.~\ref{sf} the measured effective parton distribution of the photon
and its scale dependence are presented.
The data are shown in two intervals of the parton fractional energy
a) $0.2<x_\gamma<0.4$ and b) $0.4<x_\gamma<0.7$.
The inner error bars represent the statistical errors,
the outer error bars the quadratic sum of the statistical and 
systematic errors.
In addition to the errors of the di-jet cross section, which were
described in Sect.~\ref{dijet}, the error bars include the 
uncertainty in the correction resulting from multiple parton scattering
effects which was determined using the two event generators 
PYTHIA and PHOJET (Table~\ref{pdftable}).
\begin{table}
\begin{center}
\begin{tabular}{||c|c||c|c|cr|c||}
\hhline{|t:=======:t|}
&&& \multicolumn{4}{c||}{} \\
$x_\gamma$ & \multicolumn{1}{c||}{$p_T^2[{\rm GeV}^2]$} &
 $\frac{1}{\alpha}x_\gamma \tilde{f}_\gamma(x_\gamma,p_T^2)$ &
$\sigma_{\rm stat}$
& \multicolumn{2}{c|}{$\sigma_{\rm syst}$} & $\sigma_{\rm total}$ \\
&&&& exp. & model dep. & \\
\hline\hline
0.3 & 112 &  3.11 & $\pm$ 0.07 & $\pm$ 0.82 & $\pm$ 0.31 & $\pm$ 0.88 \\
    & 224 &  4.10 & $\pm$ 0.11 & $\pm$ 1.09 & $\pm$ 0.41 & $\pm$ 1.17 \\
    & 447 &  3.91 & $\pm$ 0.25 & $\pm$ 1.25 & $\pm$ 0.48 & $\pm$ 1.34 \\
\hline
0.55 & 112 &  2.25 & $\pm$ 0.05 & $\pm$ 0.50 & $\pm$ 0.23 & $\pm$ 0.55 \\
     & 224 &  3.36 & $\pm$ 0.09 & $\pm$ 0.74 & $\pm$ 0.34 & $\pm$ 0.82 \\
     & 447 &  3.33 & $\pm$ 0.15 & $\pm$ 0.73 & $\pm$ 0.37 & $\pm$ 0.82 \\
     & 891 & 5.18 & $\pm$ 0.57 & $\pm$ 1.14 & $\pm$ 0.78 & $\pm$ 1.38 \\
\hhline{|b:=======:b|}
\end{tabular}
\end{center}
\caption{The effective photon parton distribution 
$x_\gamma \tilde{f}_\gamma$ is shown as a function of the
parton fractional energy $x_\gamma$ and the transverse momentum $p_T^2$
of the scattered partons.
The systematic errors are separated into contributions from
experimental uncertainties and model dependencies.
The total error ($\sigma_{\rm total}$) corresponds to the quadratic sum
of the statistical ($\sigma_{\rm stat}$) and systematic ($\sigma_{\rm syst}$)
errors.
\label{pdftable}}
\end{table}

For comparisons with the data, the effective parton 
distributions of the photon and the pion were calculated from the 
GRV-LO parameterizations~\cite{pgrv,pigrv}.
The pion parton distribution was 
scaled by the factor $\kappa\cdot 4\pi\alpha/f_\rho^2 \approx 0.9\,\alpha$ where
$f_\rho^2/4\pi\approx 2.2$ 
represents the probability that a photon converts into
a $\rho$ meson and
$\kappa\approx 2$ effectively accounts for contributions
of heavier vector mesons \cite{ggrv}.
This parton distribution therefore 
gives the vector meson dominance picture of the photon
and is shown as the dashed curve (VDM) in Fig.~\ref{sf}.
It differs
from the measurement both in shape and absolute rate.

The GRV-LO parameterization of the photon
parton distribution functions was chosen to resemble
that of a hadron at small scales, $\mu\approx 0.5$\,GeV (not shown).
At large scales $\mu\gg 1$\,GeV,
the parton densities of the photon reproduce 
the measurements of the photon structure function $F_2^\gamma$,
which is observed
to rise with $\mu^2=Q^2$ at a rate
which is compatible with the logarithmic increase that results from the
pointlike term in the DGLAP QCD evolution equation (\ref{dglap})
(see e.g. \cite{amy}).

%

The measurements made here show, for the first time in the context of
photoproduction, the above-mentioned logarithmic dependence of photon 
structure on the scale,
$p_T$, at which that structure is probed. 
The GRV-LO parameterization of photon structure (full curve in Fig.~\ref{sf}),
which includes the effects of the pointlike term
in the DGLAP equations,
describes the measurement.
If this term is excluded, as in the VDM model calculations shown here
(dashed curve),
it is not possible to describe the data.
Further, as information on the photon quark distributions is obtained
from the $F_2^\gamma$ measurements, it
is possible to identify the contribution they make to the
effective parton distribution measured here (dotted curve).
The difference between this and the measured values is a consequence
of the gluon content of the photon.
This is observed to
contribute about 20\% to the total effective parton distribution
in the range $0.4 < x_\gamma < 0.7$ and about 50\% in the
lower $x_\gamma$ range, $0.2 < x_\gamma < 0.4$. 

The precision of the jet measurement is similar to 
that obtained in the $e^+e^-$ experiments.
The data extend the kinematic region where the parton distributions
of the photon are measured to the scale $p_T^2=1250$\,GeV$^2$.

\section{Summary}

\noindent
A measurement of the double-differential inclusive di-jet cross section 
in terms of the parton fractional energy and the transverse energy
scale was presented from H1 photoproduction data. 
This measurement constrains the quark and gluon distributions of the photon
with a precision that is competitive with two photon data.
In addition, a new kinematic range up to $x_\gamma^{\rm jets}\approx 1$
and $(E_T^{\rm jets})^2=2500$\,GeV$^2$ is covered.

For the first time, an effective parton distribution of the 
photon was extracted from the data. 
The observed scale dependence shows an increase with $p_{T}^2$ 
that is compatible with the logarithmic increase 
predicted by perturbative QCD calculations of the 
parton content of the photon.

\subsection*{Acknowledgements}
We are very grateful to the HERA machine group whose outstanding
efforts made this experiment possible. We acknowledge the support
of the DESY technical staff. We appreciate the big effort of the
engineers and technicians who constructed and maintained the
detector. We thank the funding agencies for financial support
of this experiment. We wish to thank the DESY directorate for
the support and hospitality extended to the non--DESY members
of the collaboration.
We wish to thank M.~Drees for drawing our attention to the concept 
of the effective parton distribution and for discussions on 
the applications to the photon--proton data.
We thank M.~Klasen and G.~Kramer for enlightening discussions on the 
choice of the di-jet variables
and good collaboration.

\end{document}